\documentstyle[epsfig]{aipproc}

\begin{document}
\title{Lifetime Limit of LSP from Cosmological Light Elements\thanks{
Talk presented at cosmo-98, Asilomar, Monterey, CA, November 16--20,
1998.\protect\\
This work is supported in part by KOSEF, MOE through
BSRI 98-2468, and Korea Research Foundation.
}}

\author{Jihn E. Kim}
\address{School of Physics, Korea Institute for Advanced Study,\\
207-43 Cheongryangri-dong, Seoul 130-012, Korea, and\\
Department of Physics, Seoul National University,\\
Seoul 151-742, Korea }

\maketitle

\begin{abstract}
From the present cosmic abundance of the light elements, one can obtain
a lifetime bounds of LSP. From the consideration of deuterium, we 
obtain $\tau_{\chi} \ge 10^6$ s \cite{kkp}.
\end{abstract}

\section*{Introduction}

More than 20 years have passed since the Lee-Weinberg bound on the stable 
heavy neutrino was proposed \cite{lw}. It applies to an absolutely stable
particle which needs a conserved quantum number. The decaying particle
cosmology was propsed around the same time \cite{dkt}. But the decaying 
particle cosmology has been extensively applied for the gravitino
\cite{ekn}. In this talk, we focus on the decay of the lightest 
supersymmetric particle (LSP), $\chi$. For $\chi$ to decay, the
R-parity must be broken.

In supergravity, there can be nonrenormalizable interactions for
the R-parity violation, but we neglect these compared to the
renormalizable ones. There can be also R-violating bilinear terms,
of the form $L_iH_2$, which can be rotated away at the superpotential
level. However, in the presence of soft terms some bilinear terms
cannot be rotated away. But for decay processes, this subtle point is
not important. Thus we consider the R-violating trilinear terms,
\begin{equation}
W=\frac{1}{2}\lambda_{ijk}L^iL^jE^{ck}+\lambda^\prime_{ijk}L^iQ^jD^{ck}
+\frac{1}{2}\lambda^{\prime\prime}D^{ci}D^{cj}D^{ck}
\end{equation} 
which are allowed by (i) supersymmetry and (ii) gauge symmetry. Therefore,
it is reasonable to consider the phenomenology of R-parity violation.
With the above superpotential present, the lepton number and/or baryon
number are broken. That is the reason for requiring the
$R=(-1)^{3B+L+2S}$ conservation. Anyway, with the R-parity violation, we
expect neutrino oscillation, proton decay, and other abnormal processes.
From laboratory experiments, the single bounds on the couplings are 
not very strong, e.g. $\lambda_{121}<0.05, \lambda_{122}<0.05,
\lambda^\prime_{111}<0.001,\lambda^\prime_{112}<0.02, 
\lambda^{\prime\prime}_{112}<10^{-6}$, and $\lambda^{\prime\prime}_{113}
<10^{-5}$. However, a combined bound from lower limit of the
proton decay lifetime  is very strong
\begin{equation}
\lambda^\prime\lambda^{\prime\prime}<10^{-24}.
\end{equation} 
If a universal strength for the R-violating couplings are assumed,
then these couplings are very small. For example, if a singlet
scalar field $\phi$ carrying odd R-parity develops a VEV,
$<\phi>\equiv \epsilon M_P$, then
the R-parity is spontaneously broken. At low energy, these effects
are represented by $\lambda,\lambda^\prime,$ and $\lambda^{\prime\prime}$ 
couplings which carry negative R-parity. These may arise from 
nonrenormalizable interactions containing $\phi$ field. In this case,
these couplings can be of the same order.

Thus, to allow reasonably large R-violating couplings, one usually
fobid $\lambda^\prime$ or $\lambda^{\prime\prime}$ completely.

In general, laboratory experiments give upper bounds on the
couplings. However, the cosmological bounds depend on the region
of couplings. It is schematically shown in Fig. 1. In this talk,
I am interested in the region $\tau_\chi\sim 10^{-6}$ s.

\begin{figure}[b!] 
\centerline{\epsfig{file=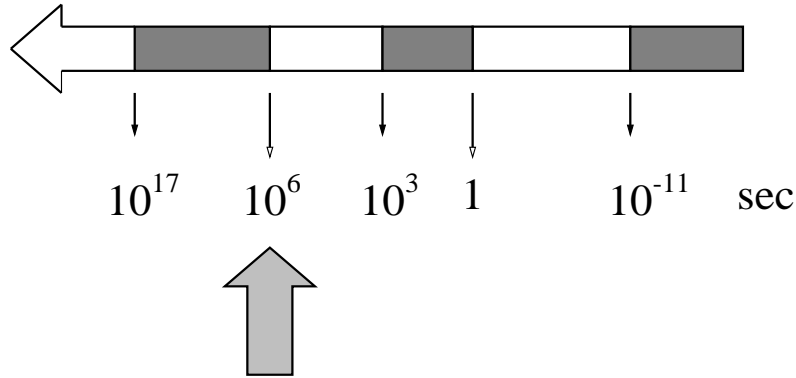,height=5cm, width=10.5cm}}
\vspace{10pt}
\caption{A schematic view of allowed windows for the $\chi$ lifetime.
Our interest is near the point marked by the grey upward arrow.}
\label{fig1}
\end{figure}

\section*{The Lifetime of LSP}

The neutralino is a mixture of the neutral gauginos and neutral
Higgsinos,
\begin{equation}
\chi=N_1\tilde B+N_2\tilde W+N_3\tilde H_1^0+N_4\tilde H_2^0.
\end{equation}
The decay can occur, for example, through $\chi\rightarrow Q+\tilde
q$ where the virtual $\tilde q$ transforms to $L+D^c$. For this
process, the interaction is given by
\begin{equation}
-\frac{\lambda^\prime_{ijk}}{M^2_{\tilde f}}\left(
gN_2+\frac{1}{3}g_1N_1\right)\int d^4\theta U^{aj}\chi
E^iD^{ck}_{a}+{\rm h.c.}
\end{equation} 
${}$For the $\chi\rightarrow u^c+d+e^+$ decay, there exist 9 diagrams.
The possible diagrams increase very rapidly if one considers colored
final states. Kinematically excluding $tt^c$ final states, there are
1,080 diagrams. Thus, we assume the {\it universal sfermion mass}
to simplify the expression. Then the $\chi$ decay rate is given by
\begin{eqnarray}
&\Gamma_{\rm tot}=\frac{g^2}{256\pi^3}\frac{m_\chi^5}{M^4_{\tilde f}}\cdot
\Large\{\sum|\lambda_{ijk}|^2\left(\frac{1}{8}N_2^2+\frac{3}{8}
N_1^2\tan^2\theta_W\right)+\sum_{j\ne 3}|\lambda^\prime_{ijk}|^2\left(
\frac{3}{4}N_2^2+\frac{7}{12}N_1^2\tan^2\theta_W\right)\nonumber\\
&+\sum_{ik}|\lambda^\prime_{i3k}|^2\left(
\frac{3}{8}N_2^2+\frac{7}{24}N_1^2\tan^2\theta_W-\frac{1}{2}
N_1N_2\tan\theta_W\right)+\sum_{k\ne 3}|\lambda^{\prime\prime}_{ijk}|^2
N_1^2\tan^2\theta_W \Large\}
\end{eqnarray}
where the overall coefficient in front of the curly bracket is
$1.8\times 10^{15}\ {\rm s}^{-1}$ for $M_{\tilde f}=1$~TeV and
$m_\chi=30$~GeV. 

\section*{Cosmological Bound on $\tau_\chi$}

\subsection*{The decoupling temperature $T_D$}

The last moment when the heavy particles are in equilibrium with
photons is the decoupling temperature $T_D$. For the neutralino,
it is determined by a competition between the universe expansion rate
and $\chi$ destruction rate. The destruction rate is a function of
sfermion mass for the $t$-channel process $\chi+\chi\rightarrow f+f^c$.
If there exists a significant Higgsino component, i.e.
$(N_1^2+N_2^2)/|N^2_{3\ {\rm or}\ 4})|>(M_Z^2/M^2_{\tilde f})^2$, then
the $s$-channel $Z$ exchange diagram dominates. It is the easiest way
to estimate the decoupling temperature. However, we also include the
sfermion exchange diagrams in the numerical estimation. Nevertheless,
the decoupling temperature does not depend on these parameters very
much. Let us introduce a decoupling factor $d$
\begin{equation}
d=\frac{m_\chi}{T_D}.
\end{equation}

The decoupling factor d is given for various parameter sets in
Ref.~\cite{kkp}. For example, $M_{\tilde f}=300$~GeV, $m_\chi=60$~GeV,
$\tan\beta=10$, and the gaugino dominated neutralino
($N_1=0.9-1$, $N_2=0.1-0.5$, and $N_3=N_4=0$) give $d=21-22$. For a wide
range of parameters, $d$ is in the range 15--30.

In particular, we note the following. Higgsino dominated $\chi$ gives 
a little bit smaller decoupling temperature and hence a lower
$\chi$ density. The $t$ and $u$ channel processes are important
for $N_{3,4}>0.1$ if $M_{\tilde f}>300$~GeV. However, the decoupling
temperature is rather insensitive to $\tan\beta$. In the numerical
analyses, we further assumed the relations $m_{A,h_1,h_2}
/m_\chi=1/3$ and $\Gamma_{A,h_1,h_2}=1/500$. We note, however, that
the cross section and the decoupling temperatures are not sensitive to 
these assumptions.  

In the next section, we are interested in relatively light neutralino
case, which excludes the possibility of producing $t,W,Z,H$ in the
$\chi\chi$ annihilation.

\subsection*{The $\chi$ lifetime bound}

If $\chi$ decays after 1 s, it affects the nucleosynthesis in a
way of destructing the already manufactured light elements.
Note that our $\chi$ lifetime is $10^{-15}/|\lambda|^2$~s and
the cosmic time scale in the radiation dominated era is
$t=0.3 N^{-1/2}M_P/T^2$. 

Requiring that $E_\chi<$~(energy
of relativistic particles), we obtain
$\tau_\chi< 4.8\times 10^{10}$~(GeV/$m_\chi)^2$~s, which gives
a $|\lambda|$ bound of order $10^{-13}$.

A more stringent bound comes from the dissociation of light elements,
in particular the D dissociation \cite{ekn,lindley}.
The $\chi$ decay products degrade very rapidly by scattering with
background photons. We are interested in the photons with 
$E>2.225$~MeV. But $e^+e^-$ production from scattering on the
background radiation turned out to be important also \cite{lindley}.
In this analysis, another important temperature parameter $T_*$
is introduced. At $T_*$, the rates for the Compton scattering and the
$e^+e^-$ production are comparable. The critical photon energy $E_*$
corresponding to $T_*$ is given by $E_*T_*= $(1/50)~MeV$^2$. Then we
note that $10^{-9}$ (with respect to the photon number) photons 
with $\omega>25T_*$ can be used to produce $e^+e^-$. In this case,
the $e^+e^-$ scatter off the background radiation to transfer the
energy to photons which can dissociate $D$. Therefore, the maximum
allowable lifetime $\tau_{\rm max}$ is determined by the condition that 
if it decayed later than $\tau_{\rm max}$, the probability for photon
to scatter with $e$ is negligible, namely it mostly scatter off with 
$D$ to dissociate it. Therefore, $\chi$ must decay before 
$\tau_{\rm max}$. The equation for $\tau_{\rm max}$ is
\begin{equation}
2\frac{n_\chi}{n_e}\frac{\epsilon}{\epsilon_0}\int\sum (\epsilon_0,T)
e^{-t/\tau_{\rm max}}\frac{dt}{\tau_{\rm max}}=1
\end{equation}
where 
\begin{equation}
\sum(\epsilon,T)=\int^{\omega_1}_{Q_D}\frac{\sigma_D}{\sigma_{KN}+
\sigma_{pp}}2\left(1-\frac{\omega}{\omega_m}\right)\frac{d\omega}{\omega_m}.
\end{equation}
Here $\epsilon$ is the photon energy in the fraction of $\chi$ mass,
$\omega$ is the photon frequency, $\sigma_D$, $\sigma_{KN}$, and
$\sigma_{pp}$ are the deuteron dissociation cross section, Klein-Nishina
cross section, and $pp$ production cross section, respectively. The
maximum scattered photon energy is $12\gamma T$. 

We calculated $\tau_{\rm max}$ from the above formula to obtain
\begin{equation}
\tau_{\rm max}=(2.2-2.5)\times 10^6\ {\rm s}.
\end{equation}  
For $\delta_B\simeq 5\times 10^{-10}$ and 60~GeV photino-like (bino-like)
neutralino, the combination of R-violating couplings is bounded as
\begin{eqnarray}
&\sum 0.12(0.05)|\lambda_{ijk}|^2
+\sum 0.31(0.07)|\lambda^\prime_{ijk}(j\ne 3)|^2
+\sum 0.04(0.04)|\lambda^\prime_{i3k}|^2\nonumber\\
&+\sum 0.23(0.12)|\lambda^{\prime\prime}_{ijk}(i<j,k\ne 3)|^2
>7.7\times 10^{-24}
\end{eqnarray} 

\section*{Conclusion}

Thus the sum of the R-violating couplings is bounded from below to
a few times $10^{-12}$, the details of which depends on the
nature of the neutralino. We studied the lifetime region around
$10^6$~s in Fig.~1. Of course, our bound is most meaningful if 
the R-violating couplings are of the same order. It is interesting
to note that there is a allowed lifetime window between $10^3$~s 
and $10^6$~s. 

\begin{figure}[b!] 
\centerline{\epsfig{file=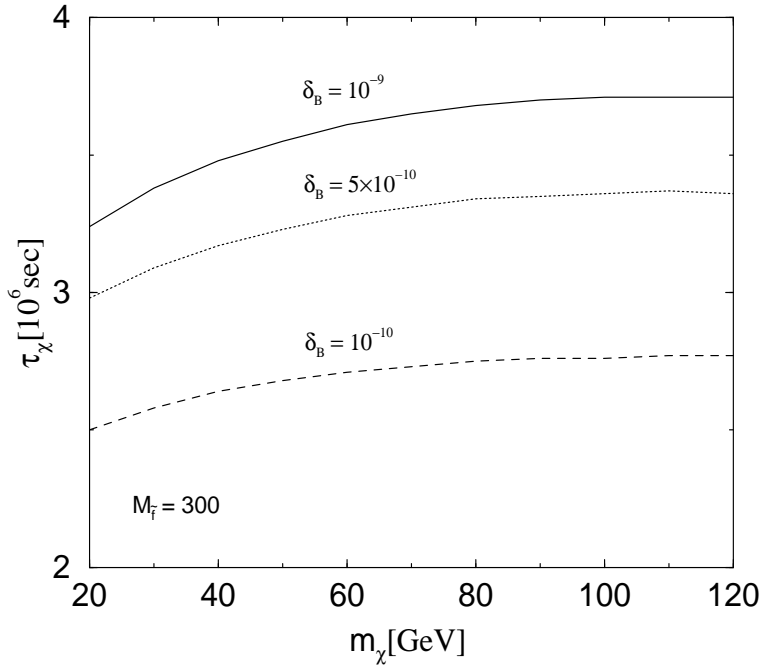,height=9cm, width=10cm}}
\vspace{10pt}
\caption{The maximum allowable lifetime of $\chi$ for $M_{\tilde f}=300$~GeV,
	$N_1=0.99$, $N_2=0.14$, and $N_3=N_4=0$.}
\label{fig2}
\end{figure}

\end{document}